\documentclass[prl,twocolumn,showpacs,epsfig,amsmath,amssymb]{revtex4}
\usepackage{amsmath}
\usepackage{graphics}
\usepackage{epsfig}
\usepackage{dcolumn}
\usepackage{bm}
\begin{document}
\title{Low-velocity anisotropic Dirac fermions on the side surface of topological 
insulators}
\author{Chang-Youn Moon, Jinhee Han, Hyungjun Lee, and Hyoung Joon Choi}
\email[Email:\ ]{h.j.choi@yonsei.ac.kr}
\affiliation{Department of Physics and IPAP, Yonsei University, Seoul 120-749, Korea}

\date{\today}

\begin{abstract}
We report anisotropic Dirac-cone surface bands on a side-surface geometry 
of the topological insulator Bi$_2$Se$_3$ revealed by first-principles 
density-functional calculations. We find that the electron velocity in the 
side-surface Dirac cone is anisotropically reduced from that in the 
(111)-surface Dirac cone, and the velocity is not in parallel with the 
wave vector {\bf k} except for {\bf k} in high-symmetry directions. 
The size of the electron spin depends on the direction of {\bf k} due to 
anisotropic variation of the noncollinearity of the electron state.
Low-energy effective Hamiltonian is proposed for side-surface Dirac
fermions, and its implications are presented including refractive
transport phenomena occurring at the edges of tological insulators where 
different surfaces meet.
\end{abstract}

\pacs{71.15.Mb, 73.20.-r, 73.61.Ng, 75.70.Tj}

\maketitle
Topological insulators (TIs) are characterized by odd number of 
Dirac-cone-like surface bands which guarantees robust metallicity
against arbitrary perturbations preserving time-reversal 
symmetry \cite{Fu2007,Moore}, and by the helical spin 
texture of the surface states which leads to remarkable transport 
properties such as the absence of backscattering by 
non-magnetic defects as confirmed by experiments
\cite{Roushan,T_Zhang,Alpichshev}. The latter feature reminds us 
of graphene where charge carriers are also described as massless 
Dirac fermions,
as demonstrated intriguingly by 
the Klein tunneling \cite{Klein} where normal
incident wave exhibits perfect transmission through any finite 
potential barrier. A crucial difference is that TI involves 
real spins whereas graphene does pseudo-spins, so that, for 
example, one-dimensional wires of TI have maximum 
magnetoconductance at {\it half integral} flux quanta due to $\pi$ 
Berry phase acquired when circling around the wires
\cite{Bardarson,Vishwanath}.

Promising TI materials \cite{Zhang,Lin} include series of 
rhombohedral materials Bi$_2$Se$_3$, 
Bi$_2$Te$_3$, and Sb$_2$Te$_3$ proposed by a recent first-principles 
calculation \cite{Zhang}, with the existence of single Dirac cone on 
their (111) surfaces indeed confirmed by angle-resolved photoemission 
spectroscopies \cite{Xia,Hsieh}. Among them, Bi$_2$Se$_3$ is considered 
most promising in the application point of view due to its relatively 
large bulk band gap of $\sim$300~meV that can evade bulk conduction 
of thermally excited charge carriers at room temperature. This material has a 
layered structure where the Bi$_2$Se$_3$ quintuple layers (QLs) are stacked 
along the rhombohedral (111) direction \cite{Wyckoff} combined by van der Waals 
type interaction. Thus, a natural cleaving face is the (111) surface
between two QLs, and it has been the only surface 
considered theoretically or experimentally so far \cite{Song}.

Surfaces other than the (111) surface, and interfaces between 
different surfaces, appear inevitably as step edges on cleaved (111) surfaces
\cite{T_Zhang,Alpichshev,Seo} and as side surfaces in
nanowires or nanoribbons \cite{Kong,Cha,Hong,Peng}.
According to the topological band theory \cite{Fu2007,Moore}, 
a strong TI like Bi$_2$Se$_3$ \cite{Zhang} should have odd 
number of Dirac cones on any arbitrary surface.  However, existence of 
Dirac cones on a surface other than the (111) surface and their
physical properties have never been studied by realistic electronic-structure 
calculations or by direct measurements.

In this Letter, we investigate the electronic structure of a side surface 
of Bi$_2$Se$_3$ using first-principles density-functional calculations,
and find a Dirac-cone band structure, for the first time on a 
surface other than the (111) surface, consistently with the $Z_2$ 
topological band theory. We find that the side-surface Dirac cone 
has lower symmetry, with elliptic equi-energy contour lines,
and smaller velocity, introducing the refractive index
in analogy with optics.
We also present the low-energy Hamiltonian for the side-surface Dirac cone,
and reflection and transmission properties at interfaces 
between different surfaces.

\begin{figure}
\centering
\epsfig{file=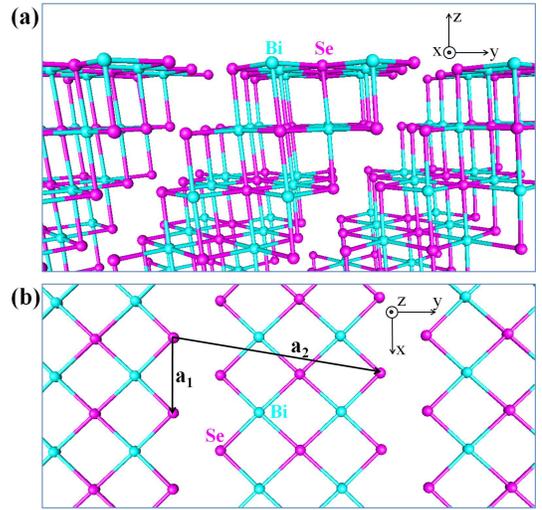,width=7cm,angle=0,clip=}
\caption{(Color online). Atomic structure of the Bi$_2$Se$_3$ (221)
surface. (a) Side view with perspective, showing slanted QLs.
(b) Top view of the topmost atomic layer, with
surface unit vectors denoted by {\bf a}$_1$ and {\bf a}$_2$.
Using the experimental bulk structure \cite{Wyckoff},
the length of {\bf a}$_1$ is 4.14~{\AA}, the $y$-component of {\bf a}$_2$
is 11.26~{\AA} long, and the QLs are slanted by 58$^\circ$.
}
\label{fig1}
\end{figure}

Our first-principles calculations are based on the density-functional 
theory with the generalized gradient approximation \cite{PBE} and 
{\it ab-initio} norm-conserving pseudopotentials as implemented 
in SIESTA code \cite{SIESTA}. We employ partial-core correction for Bi 
and Se pseudopotentials, pseudo-atomic orbitals 
(double-$\zeta$-polarization basis set) for electronic wavefunctions, 
and a cutoff energy of 500 Ry for real-space mesh. 
We use a supercell containing a slab of 
Bi$_2$Se$_3$ and 9 {\AA} thick vacuum to study surface states while preventing
interaction between periodic images. Brillouin-zone (BZ) 
integration is performed with a two-dimensional (2D) 
12 $\times$ 6 $k$-point grid. 
Spin-orbit interaction is treated by relativistic pseudopotentials 
in the fully nonlocal form \cite{Kleinman}. 
Atomic positions are taken from the experimental bulk structure \cite{Wyckoff}.

Figure~1 shows the atomic structure of the side surface considered
in our present work. This side surface is stoichiometric in the
sense that the topmost atomic layer of the surface and other atomic 
layers in parallel with it have stoichiometric Bi and Se compositions
for Bi$_2$Se$_3$ within each atomic layer. The normal direction of 
the side surface is (221) in Miller indices of the rhombohedral bulk
structure, making an angle of 58$^\circ$ to the (111) plane. 
In Fig.~1, no bond is drawn 
between Se atoms belonging to different QLs to represent the (111) 
cleaving planes with real-space gaps which are along the $x$ direction 
and repeated in the $y$ direction. The (221) surface has lower symmetry
than the (111) surface, exhibiting anisotropy between $x$ and $y$ 
directions clearly. 
We construct a slab of eleven atomic layers of Bi$_2$Se$_3$ 
with the slab thickness of $\sim$30 {\AA} 
to study surface electronic structures.

\begin{figure}
\centering
\epsfig{file=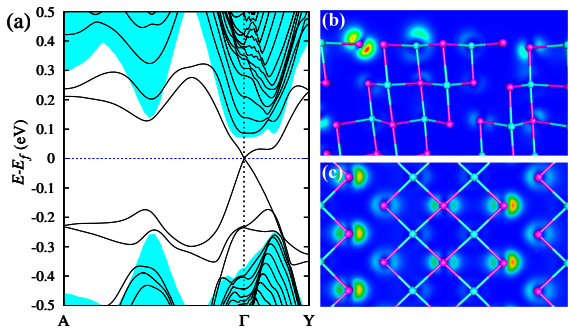,width=7.5cm,angle=0,clip=}
\caption{(Color online). Electronic structure of the Bi$_2$Se$_3$ (221) surface.
(a) Calculated side-surface electronic band
structure from the BZ center to BZ boundaries along the two
reciprocal lattice vectors. Filled regions denote bulk continuum states
projected on the surface BZ. (b) Side view and (c) top view of a wavefunction
at the Dirac point,  which show that the state is localized near the surface.
}
\label{fig2} 
\end{figure}

Figure 2(a) shows calculated electronic band structure along two lines in 
2D BZ: from $\Gamma$ to $A$ and from $\Gamma$ to $Y$. Here, $A$ and $Y$ 
are BZ boundaries along the reciprocal lattice vectors {\bf b}$_1$ and 
{\bf b}$_2$, respectively, associated with the real-space unit 
vectors {\bf a}$_1$ and {\bf a}$_2$ shown in
Fig.~1(b). In Fig.~2(a), a crossing of linear bands is clearly visible 
at the $\Gamma$ point with the Dirac-point energy coinciding with the Fermi
energy ($E_f$). The bulk band continuum is projected
onto the 2D surface BZ as filled areas in Fig.~2(a), where 
the bulk band gap is about 300~meV, consistently with previous
calculation \cite{Zhang}. To check if the crossing bands are indeed 
surface states, we plot the squared amplitude of the wavefunction,
$|\psi({\bf r})|^2$, for a state at the $\Gamma$ point in Figs.~2(b) and (c). 
We find that it is mainly located on the surface atoms and decays 
rapidly into the bulk region, showing the surface-state nature of the 
crossing bands. Around the $\Gamma$ point in Fig. 2(a), we notice that the 
slope of the linear band is steeper, i.e. the Fermi velocity is larger, 
along the $\Gamma$--$A$ direction than $\Gamma$--$Y$, indicating
anisotropy of the Dirac cone. Particle-hole asymmetry is also visible,
where band linearity persists down to $\sim$200 meV {\it below} 
$E_f$ while the two bands bend significantly only $\sim$30 meV 
{\it above} $E_f$.

\begin{figure}
\centering
\epsfig{file=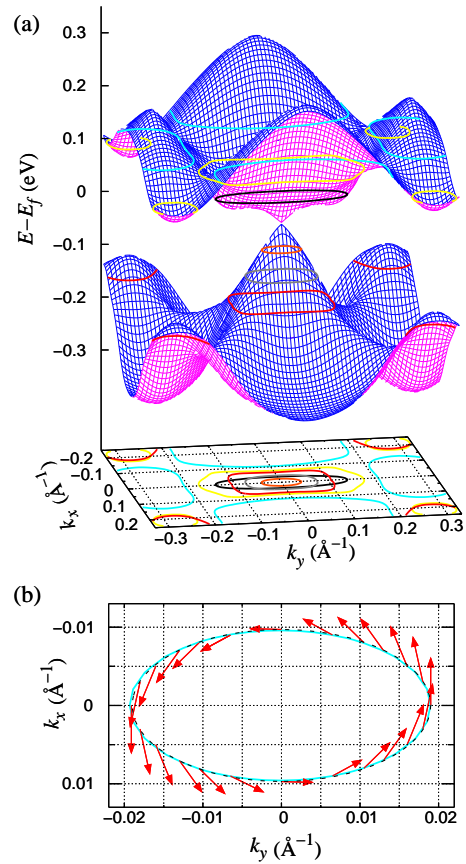,width=6cm,angle=0,clip=}
\caption{(Color online). The Dirac cone on the Bi$_2$Se$_3$ (221) surface.
(a) Calculated 2D electronic band structure
near the Dirac point at {\bf k} = 0. Equi-energy contours for
-150, -100, -50, 50, 100, and 150~meV from the Dirac point
are drawn on the energy surface and then projected onto the bottom plane.
(b) Equi-energy contour at 20 meV below the Dirac point.
Arrows denote calculated spins of the states at each ${\bf k}$.
}
\label{fig3}
\end{figure}

The shape of the side-surface Dirac cone is more obvious in the 2D
plot of the bands near the $\Gamma$ point, as shown in Fig.~3(a). 
The Dirac cone is anisotropic, with the slope of the band and hence 
the Fermi velocity being larger in the $x$ direction than in the $y$ 
direction. As mentioned above, the linear band feature holds only within 
a very small energy window near $E_f$, so the energy contours of 
50~meV above $E_f$ (in black ) and below $E_f$ (in orange) look very 
different from each other [Fig.~3(a)].
At energies near $E_f$ close enough to ensure the band 
linearity, the energy contours are basically very close to ellipses. 

We obtain the spin of a Dirac-cone state on the (221) surface by evaluating 
expectation values of the three spin-operator components, and the results 
are shown
in Fig.~3(b). As expected for the helical spin fermion, 
the spin is almost perpendicular to the wave vector ${\bf k}$. Meanwhile, 
since each electronic state in this material is not a spin eigenstate but 
a noncollinear spin state due to the spin-orbit interaction, the spin size 
is not necessarily 1 $\mu_B$ in general \cite{Louie}. Unlike the (111) surface  where 
the spin size is independent of {\bf k}, the spin size in the
(221)-surface Dirac cone is found to depend on {\bf k}, having its maximal value
of 0.83~$\mu_B$ at $k_x=0$ and gradually decreasing toward 0.55~$\mu_B$ 
at $k_y=0$ around the Dirac cone [Fig.~3(b)].

Now we consider low-energy effective
Hamiltonian similar to the previously suggested one \cite{Zhang} which
basically corresponds to a Rashba-type surface spin-orbit Hamiltonian. 
Considering the reduced symmetry, the 
(221)-surface Hamiltonian in the first order of $k$ would be 
\begin{equation}
H=\hbar(-v_x\sigma_yk_x + v_y\sigma_xk_y),
\end{equation}
where $\sigma_x$ and $\sigma_y$ are the Pauli matrices. 
The Fermi velocities $v_x$ and $v_y$ in $x$ and $y$ directions
on the (221) surface
are $3.1\times10^5$ m/s and $1.4\times10^5$ m/s, respectively, from our 
first-principles calculations. 
Energy eigenvalues are 
\begin{equation}
E=\pm \hbar\sqrt{v_x^2 k_x^2 + v_y^2 k_y^2},
\end{equation}
and the corresponding eigenstates are 
\begin{equation}
\psi_{\bf k}({\bf r}) = \frac{
{\rm e}^{i{\bf k}\cdot{\bf r}}}{\sqrt{2}}
\!\left(\!\!\begin{array}{c}1\\ \pm (-i v_x k_x \!+\!  v_y k_y)/
\!\sqrt{v_x^2 k_x^2 \!+\!v_y^2 k_y^2}\end{array}\!\right).
\end{equation}
In Eqs.~(2) and (3), the plus sign is for the conduction band while 
the minus sign is for the valence band.
The equi-energy contour from Eq.~(2) is an
ellipse, which agrees with Fig.~3(b) at $E$ = $-$20 meV.

As mentioned earlier, interfaces of different surfaces are very often 
involved in many experimental situations where two different surfaces 
meet as a form of step edges on a cleaved TI surface or geometric edges 
in TI nanowires and nanoribbons. As our result shows that different 
TI surfaces have Dirac cones with different velocities, we can expect 
reflection and refraction behaviors in electronic transports at the
interfaces of TI surfaces in close analogy with optics
with distinct refractive-index materials. 
The concept of the refractive index can be introduced to a TI surface,
and it provides a new insight for electronic transport in various TI 
geometries. We discuss two basic cases, a simple interface of TI 
surfaces and a Fabry-Perot interferometer.

\begin{figure}
\centering
\epsfig{file=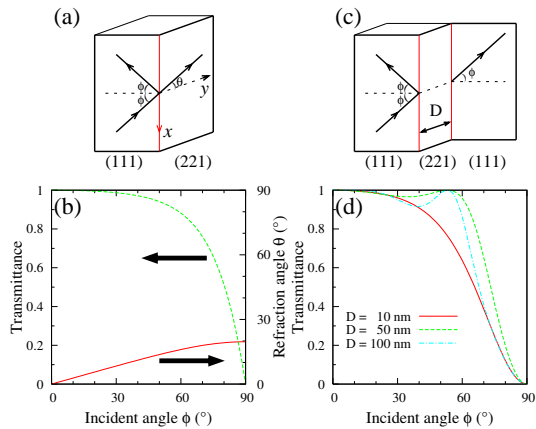,width=7.0cm,angle=0,clip=}
\caption{(Color online).
Ray optics of Dirac fermions in TI.
(a) Schematic diagram of the edge of the (111) and (221) surfaces
in Bi$_2$Se$_3$.
(b) Transmission probability $T$ and refraction
angle $\theta$ as functions of incident angle $\phi$ in (b).
(c) Schematic diagram of the Fabry-Perot interferometer consisting of
(111)/(221)/(111) surfaces.
(d) $T$ as a function of $\phi$ with the (221)-surface widths
of 10, 50, and 100 nm in (c).
}
\label{fig4}
\end{figure}

First we consider a simple interface between (111) and (221) surfaces
which is parallel to the $x$ direction [Fig.~4(a)].
When a Dirac-cone state of the wave vector {\bf k} is incident from
the (111) surface to the interface at $y=0$,
with the incident angle $\phi$, we can obtain
the transmission probability $T$ and the refraction angle $\theta$ 
from the equality of $x$-components of incident, 
reflected, and transmitted wave vectors, the energy conservation, and the continuity of the wave
functions at the interface, similarly to the graphene case previously
studied \cite{Neto,Klein,Park}: 
\begin{equation}
T=\frac{{\rm cos}(\phi-\theta') + {\rm cos}(\phi+\theta')}{1+ {\rm cos}
(\phi+\theta')},
\end{equation}
where $\theta' \equiv {\rm tan^{-1}}(v_xk_x/v_yq_y)$ 
appears from the relative phase of the two spinor components in 
Eq.~(3). The refraction angle $\theta$ is given 
by $\theta={\rm tan^{-1}}(k_x/q_y)$, 
where $k_x=k~{\rm sin}\phi$ and 
$q_y=\sqrt{v_0^2 k^2\!-\!v_x^2 k_x^2}/v_y$
is the $y$-component of the wave vector in the (221) surface region.
The Fermi velocity $v_0$ in the (111) surface is about 
5~$\times$~10$^5$~m/s in our calculation, consistently with previous 
work \cite{Zhang}. 
Here we note that the incident and 
refraction angles satisfy the Snell's law with the refractive index $n$ 
of the TI surfaces defined as the inverse of the phase velocity
($v_{ph} = E/\hbar k$) of Dirac fermions, 
which, in general, depends on 
the {\bf k} direction but not on the size of {\bf k}.

Calculated transmission probability $T$ and refraction angle $\theta$
are plotted in Fig.~4(b) as functions of incident angle $\phi$ at
$E$ = 20 meV. At normal incidence ($\phi=0$), $T$ is 1, which 
means a perfect transmission, demonstrating the absence of back scattering
due to the spin texture. As $\phi$ increases to 90$^\circ$, $T$ slowly
decreases to 0, being greater than 0.5 for $\phi$ less than
80$^\circ$. At the
same time, the refraction angle $\theta$ increases slowly from 0$^\circ$ 
and reaches 20$^\circ$ at $\phi=90^\circ$. Thus, in the opposite
situation in which the Dirac fermion is incident from the side surface to 
the (111) surface, the refraction angle reaches 
$90^\circ$ already at the incident angle of 20$^\circ$ and at larger 
incident angles the Dirac fermion cannot transmit through the interface.
This implies that {\it the total internal reflection} can occur when the 
electron is incident from 
the side surface to the (111) surface, 
similarly to the optical system where the incident wave 
resides in a dielectric with higher refractive index than the other side 
of the interface. This opens a chance to produce evanescent waves
of Dirac fermions on the (111) surface and their related phenomena.

Another relevant geometry is a Fabry-Perot interferometer where
the (221) surface is made between two (111) surfaces [Fig.~4(c)], 
which might be a possible candidate to describe the step edges on the 
(111) surface in many previous experiments \cite{T_Zhang,Alpichshev}.
With boundary conditions similar to the above interface problem, we obtain
reflectivity $r$ and hence the reflectance $R=|r|^2$
in analogy with the graphene case \cite{Klein} 
\begin{equation}
R  =  \frac{{\rm sin}^2(q_y D)({\rm sin}\phi-{\rm sin}\theta')^2}{
{\rm cos}^2\phi~{\rm cos}^2\theta'
 +{\rm sin}^2(q_yD)({\rm sin}\phi-{\rm sin}\theta')^2},
\end{equation}
where 
$D$ is the width of the 
(221) surface region [Fig.~4(c)], 
and the other parameters are defined similarly to the simple-interface 
case above. 

The transmission probability through the interferometer,
$T=1-R$, is plotted as a function of the incident angle 
$\phi$ in Fig.~4(d), considering three different side-surface widths $D$ =  
10, 50, and 100 nm, with $E$ =  20 meV. For all cases,
we have the perfect transmission for the normal incidence ($\phi=0$) as 
in the simple-interface case, because of the forbidden back 
scattering. We also expect a perfect transmission to occur resonantly when the 
destructive interference happens between the incident and 
reflected waves, i.e., $q_yD=n\pi$ for positive integer $n$ from Eq.~(5). 
Because of relatively low Fermi velocity in the side surface region, 
$D$ = 10 nm is not large enough for the resonant transmission to occur 
at $E$ =  20 meV, so $T$ monotonically decreases from 1 to 0 with 
increasing $\phi$. For larger widths, 50 and 100 nm, $T$ reaches 1 at 
$\phi = 53^\circ$, where the 
resonance condition is fulfilled. 
In the case of $D$ = 100 nm, the resonance at $\phi = 53^\circ$
is narrower, with less transmission at $\phi$ between 0$^\circ$ and 
53$^\circ$. 

Our Fabry-Perot structure can also be considered as an 
electronic analogy of the `optical fiber' where propagating waves are 
confined within the high refractive index (low Fermi velocity) region 
by the total internal reflection. If the side surface region is clean 
enough to prevent the waves from escaping into the (111) surface region
with high incident angles by defect scattering, this structure might
be used as a one-dimensional conducting wire.

In conclusion, we performed first-principles calculations to 
investigate the electronic structure of a side surface of 
topological insulator Bi$_2$Se$_3$, finding a Dirac cone 
on the side surface, consistently with the topological-band-theory 
argument. The side-surface Dirac cone is found to be slow and
anisotropic in an elliptic shape, with the spin size dependent on
{\bf k}. The low-energy effective Hamiltonian is proposed for 
the side-surface Dirac fermions, and the concept of the
refractive index is introduced to the TI surface. Refraction, reflection,
and transmission behaviors of Dirac fermions
are demonstrated in the simple TI interface and
the Fabry-Perot TI interferometer. Our idea of making TI junctions
using different surfaces can be applied to form more complex
structures such as superlattices, opening a chance for novel physical
properties of Dirac fermions.

\begin{acknowledgments}
We acknowledge helpful discussions with Jung Hoon Han
and Young-Woo Son. This work was supported by NRF of Korea 
(Grant No. 2009-0081204) and KISTI Supercomputing Center 
(Project No. KSC-2008-S02-0004). 

\end{acknowledgments}


\end{document}